\documentclass{PoS}
\usepackage[utf8x]{inputenc}
\usepackage{mciteplus}

\newcommand{\xt}{{\mathbf{x}_T}}

\newcommand{\yt}{{\mathbf{y}_T}}

\newcommand{\tr}{\, \mathrm{Tr} \, }

\newcommand{\nc}{{N_\mathrm{c}}}

\newcommand{\gev}{\ \textrm{GeV}}

\newcommand{\as}{\alpha_{\mathrm{s}}}

\title{Balitsky-Kovchegov equation at next-to-leading order accuracy with a resummation of large logarithms}

\ShortTitle{DIS16}

\author{T. Lappi\\
Department of Physics, University of Jyväskylä, %
 P.O. Box 35, 40014 University of Jyv\"askyl\"a, Finland\\
 and \\
 Helsinki Institute of Physics, P.O. Box 64, 00014 University of Helsinki,
Finland \\
        E-mail: \email{tuomas.v.v.lappi@jyu.fi}}

\author{\speaker{H. Mäntysaari}\\
        Physics Department, Brookhaven National Laboratory, Upton, NY 11973, USA\\
        E-mail: \email{mantysaari@bnl.gov}}

\abstract{We include resummation of large transverse logarithms into the next-to-leading order Balitsky-Kovchegov equation. The resummed NLO evolution equation is shown to be stable, the evolution speed being significantly reduced by higher order corrections. The contributions from $\as^2$ terms that are not enhanced by large logarithms are found to be numerically important close to phenomenologically relevant initial conditions.}

\FullConference{XXIV International Workshop on Deep-Inelastic Scattering and Related Subjects\\
		11-15 April, 2016\\
		DESY Hamburg, Germany}

\begin{document}

\section{Introduction}
Proton structure function measurements at HERA have shown that the gluon density of the proton grows rapidly at small-$x$~\cite{Aaron:2009aa,Abramowicz:2015mha}. This phenomenon can be understand from perturbative QCD, as emission of gluons carrying small momentum fraction is favored. At sufficiently small $x$, the gluon densities are then expected to become so large that one has to also take into account non-linear phenomena such as gluon recombination. This high-density gluonic matter can be described within the Color Glass Condensate (CGC) framework (for a review, see e.g. Ref.~\cite{Gelis:2010nm}).

In the CGC, the Bjorken-$x$ evolution of the hadron wavefunction is given by the evolution equations such as the Balitsky-Kovchegov (BK)~\cite{Balitsky:1995ub,Kovchegov:1999yj} or JIMWLK~\cite{JalilianMarian:1996xn,JalilianMarian:1997jx, JalilianMarian:1997gr,Iancu:2001md, Ferreiro:2001qy, Iancu:2001ad, Iancu:2000hn} equations. The leading order CGC calculations (with running coupling corrections) have been successful in describing many collider experiments, such as Deep Inelastic Scattering~\cite{Albacete:2010sy,Rezaeian:2012ji} and single~\cite{Albacete:2010bs,Tribedy:2011aa,Rezaeian:2012ye,Lappi:2013zma,Ducloue:2015gfa} and double inclusive particle production~\cite{Albacete:2010pg,Stasto:2011ru,Lappi:2012nh,JalilianMarian:2012bd}. 
%The CGC framework has also been successfully applied to calculations of the initial state for hydrodynamical modeling of a heavy ion 
%collision~\cite{Lappi:2011ju,Schenke:2012wb,Gale:2012rq} and exclusive vector meson production\cite{Kowalski:2006hc,Mantysaari:2016ykx}.

In the phenomenological CGC calculations the necessary ingredients are the dipole amplitude at initial Bjorken-$x$ (non-perturbative input, obtained by fitting the HERA DIS data~\cite{Albacete:2010sy,Rezaeian:2012ji,Lappi:2013zma}), the evolution equation for the dipole amplitude in $x$ (e.g. the BK equation) and the cross section for the particular process. In recent years, first steps beyond the leading order accuracy have been taken by deriving the BK~\cite{Balitsky:2013fea} and JIMWLK evolution equations \cite{Kovner:2013ona} and some cross sections~\cite{Altinoluk:2011qy,Chirilli:2011km,Chirilli:2012jd,Stasto:2013cha,Altinoluk:2014eka,Ducloue:2016shw} at next-to-leading order accuracy (see also Ref.~\cite{Lappi:2012vw} for a proposal to include running coupling corrections to JIMWLK).

The next-to-leading order BK equation~\cite{Balitsky:2008zza} was solved numerically for the first time in Ref.~\cite{Lappi:2015fma}, where it was shown that phenomenologically relevant initial conditions yield to unphysical solutions. In particular, the equation was shown to give negative evolution speed which would correspond do decrease of unintegrated gluon distribution when Bjorken-$x$ is decreased. The origin of this problem was traced back to so called non-conformal double logarithm~\cite{Balitsky:2009xg} modifying the leading order part of the BK kernel. Later, a resummation scheme to resum large logarithmic contributions to all orders has been developed~\cite{Iancu:2015vea,Iancu:2015joa}. The first solution to the NLO BK equation improved with these resummations was presented in Ref.~\cite{Lappi:2016fmu}.

\section{The NLO BK equation}
The next to leading order BK equation with resummation of large single and double logarithmic corrections was written in Ref.~\cite{Lappi:2016fmu} in the large-$\nc$ and mean field limit. The equation reads
\begin{equation}
\partial_y S(r) = \frac{\as(r)\nc}{2\pi^2} \left[K_\mathrm{Bal} K_\mathrm{DLA} K_\mathrm{STL}  - K_\mathrm{sub} + K_1^\mathrm{fin}\right] \otimes D_1 + \frac{\as^2 \nc^2}{8\pi^4} K_2 \otimes D_2 + \frac{\as^2 n_f N_c}{8\pi^4} K_f \otimes D_f.
\end{equation}
Here $\otimes$ refers to integral over the transverse position of the emitted gluon (in case of $D_1$) or both of the emitted gluons ($D_2$ and $D_f$). Here, $D_1$, $D_2$ and $D_f$ are functions of the correlator of two Wilson lines, the dipole operator $S = 1/\nc \langle \tr U(\xt)U^\dagger(\yt)\rangle$. The Balitsky running coupling kernel~\cite{Balitsky:2006wa} is denoted by $K_\mathrm{Bal}$, and $K_\mathrm{DLA}$ resums double logarithmic corrections and removes the double log term that caused the dipole amplitude to have unphysical evolution in Ref.~\cite{Lappi:2015fma}. For explicit expressions, we refer the reader to Ref.~\cite{Lappi:2016fmu}.

% Large single transverse logarithms are resummed by $K_\mathrm{STL}$, which also includes the same contribution (in leading log accuracy) that is written in the order $\as^2$ in $K_2$. This double counting is removed by introducing $K_\mathrm{sub}$. For explicit expressions, we refer the reader to Ref.~\cite{Lappi:2016fmu}.

The resummation of single transverse logarithms done by $K_\mathrm{STL}$ is derived in Ref.~\cite{Iancu:2015joa}. The resummation is done at leading log accuracy, 
and the $\as^2$ contribution of the resummed logarithms 
is included exactly in the kernel $K_2$. This double counting is removed by subtracting $\as^2$ part of the resummation factor $K_\mathrm{STL}$, denoted by $K_\mathrm{sub}$.

As the resummation is done at leading log accuracy, it does not fix the numerical value of the constant factor $C_\mathrm{sub}$ (which should be of the order one) in
\begin{equation}
K_\mathrm{STL} = \exp \left\{ -\frac{\as \nc A_1}{\pi} \left| \ln \frac{C_\mathrm{sub} r^2}{\min\{X^2,Y^2\}} \right| \right\},
\end{equation}
where $A_1=11/12$ and $X$ and $Y$ are the sizes of the two daughter dipoles formed in the emission of a gluon from the parent dipole. We fix the value of $C_\mathrm{sub}$ by requiring that $K_\mathrm{sub}$ reproduces as accurately as possible the small-$r$ limit of the other NLO terms. This we find to happen when $C_\mathrm{sub}=0.65$, and with this choice the largest possible amount of NLO contributions are included in the resummation contribution which is numerically  easier to calculate.

\section{Evoution of the dipole amplitude}

\begin{figure}
\begin{minipage}[t]{0.48\linewidth}
\centering
\includegraphics[width=1.05\textwidth]{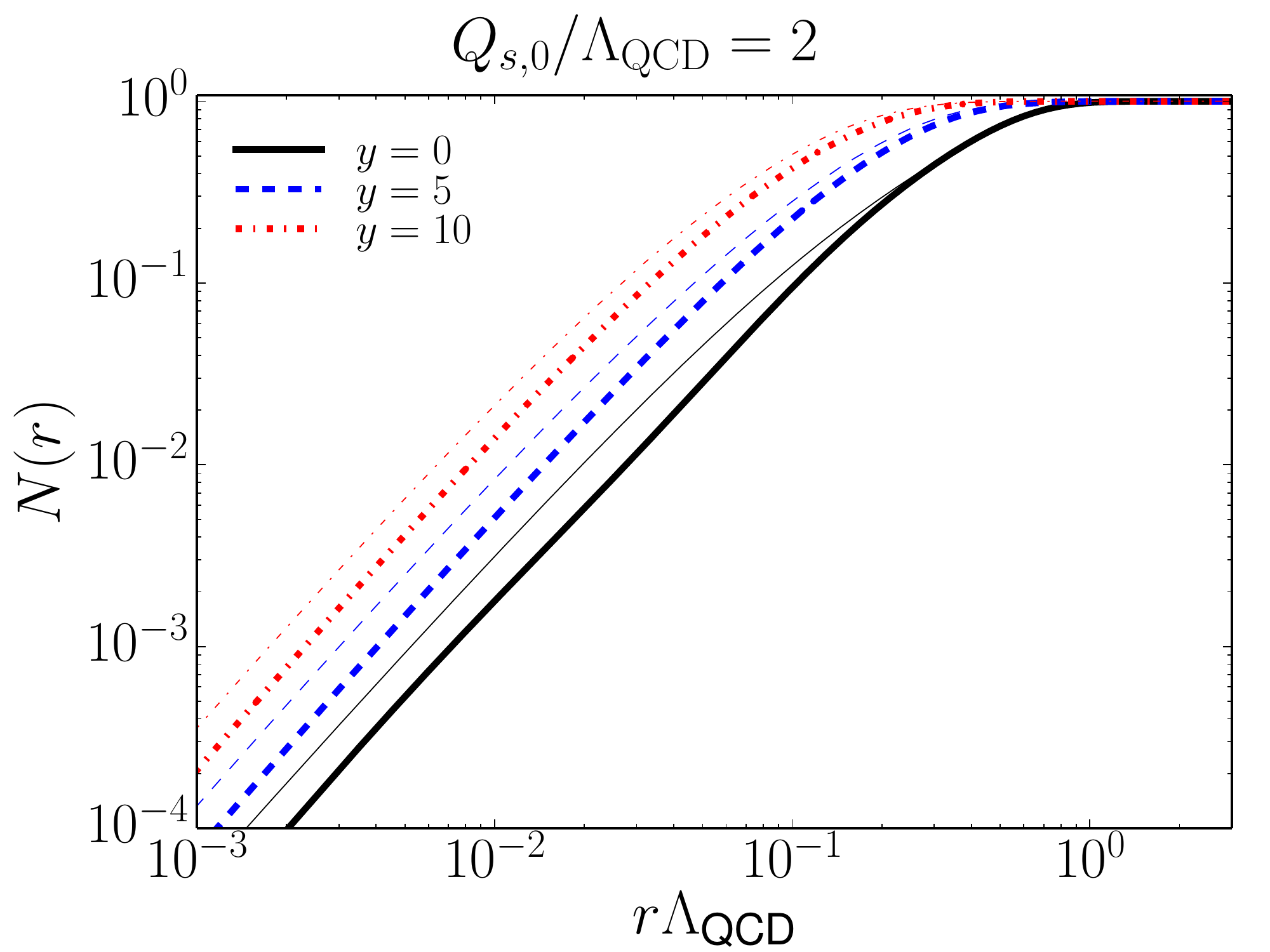} 
\caption{Dipole amplitude at the initial condition and after $5$ and $10$ units of rapidity evolution. For comparison, the corresponding amplitudes obtained by using MV model without resummation as an initial condition are shown as thin lines.}
\label{fig:amplitude} 
\end{minipage}
\hspace{0.5cm}
\begin{minipage}[t]{0.48\linewidth}
\centering
\includegraphics[width=1.05\textwidth]{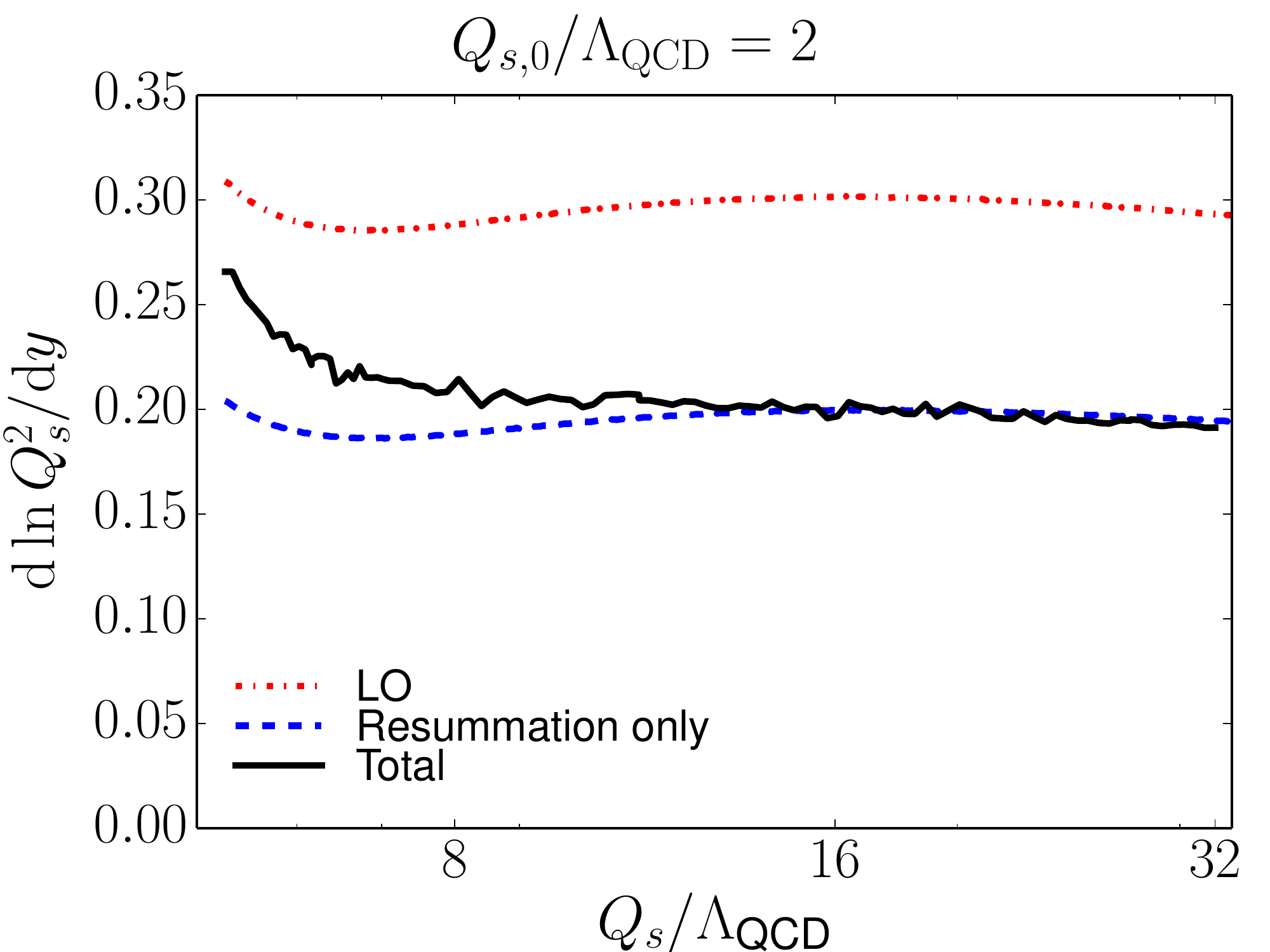} 
\caption{Evolution speed of the saturation scale solved from the leading order BK equation with running coupling (dashed-dotted line), LO equation improved with resummation of large transverse logarithms (dashed line) and the full NLO BK equation with resummation corrections (solid line).}
\label{fig:evolutionspeed} 
\end{minipage}
\end{figure}

We solve numerically the NLO BK equation improved by including resummations of large logarithms. The initial condition is the MV model with resummation corrections introduced in Ref.~\cite{Iancu:2015vea}. For comparison, we also solve the equation by using the same initial condition without resumming it. The obtained dipole amplitudes $N(r)=1-S(r)$ as a function of dipole size are shown in Fig.~\ref{fig:amplitude}, where the amplitude is shown after $5$ and $10$ units of rapidity evolution. We find that, unlike the case of the NLO BK equation without resummation, one obtains a stable evolution toward larger rapidities. Resummation has large effect on the initial condition, but the evolution at larger rapidities does not depend on the details of the initial condition.

The evolution speed is studied in more detail in Fig.~\ref{fig:evolutionspeed}, where the rapidity derivative of the saturation scale is shown. The saturation scale $Q_s^2$ is defined as
\begin{equation}
 	N(r^2=2/Q_s^2) = 1 - e^{-1/2}.
\end{equation}
The resummed NLO BK equation, labeled as \emph{Total} in Fig.~\ref{fig:evolutionspeed} is found to evolve significantly (about $30\%$) slower than the leading order equation that has running coupling corrections included. This is expected, as the leading order fits to HERA DIS data tend to prefer slower evolution speeds than what one would naturally obtain within the CGC picture~\cite{Lappi:2013zma}. At large saturation scales we find that the fixed order $\as^2$ corrections become negligible, as the evolution speeds obtained from full resummed NLO BK equation and leading order equation with resummations are approximately the same. However, close to initial condition which is expected to be in the phenomenologically relevant range, having a saturation scale $\sim 1 \gev$, the contribution from fixed order $\as^2$ terms are large. They are numerically much more demanding to compute, but our results suggest that they should not be neglected.

\begin{figure}
\begin{minipage}[t]{0.48\linewidth}
\centering
\includegraphics[width=1.05\textwidth]{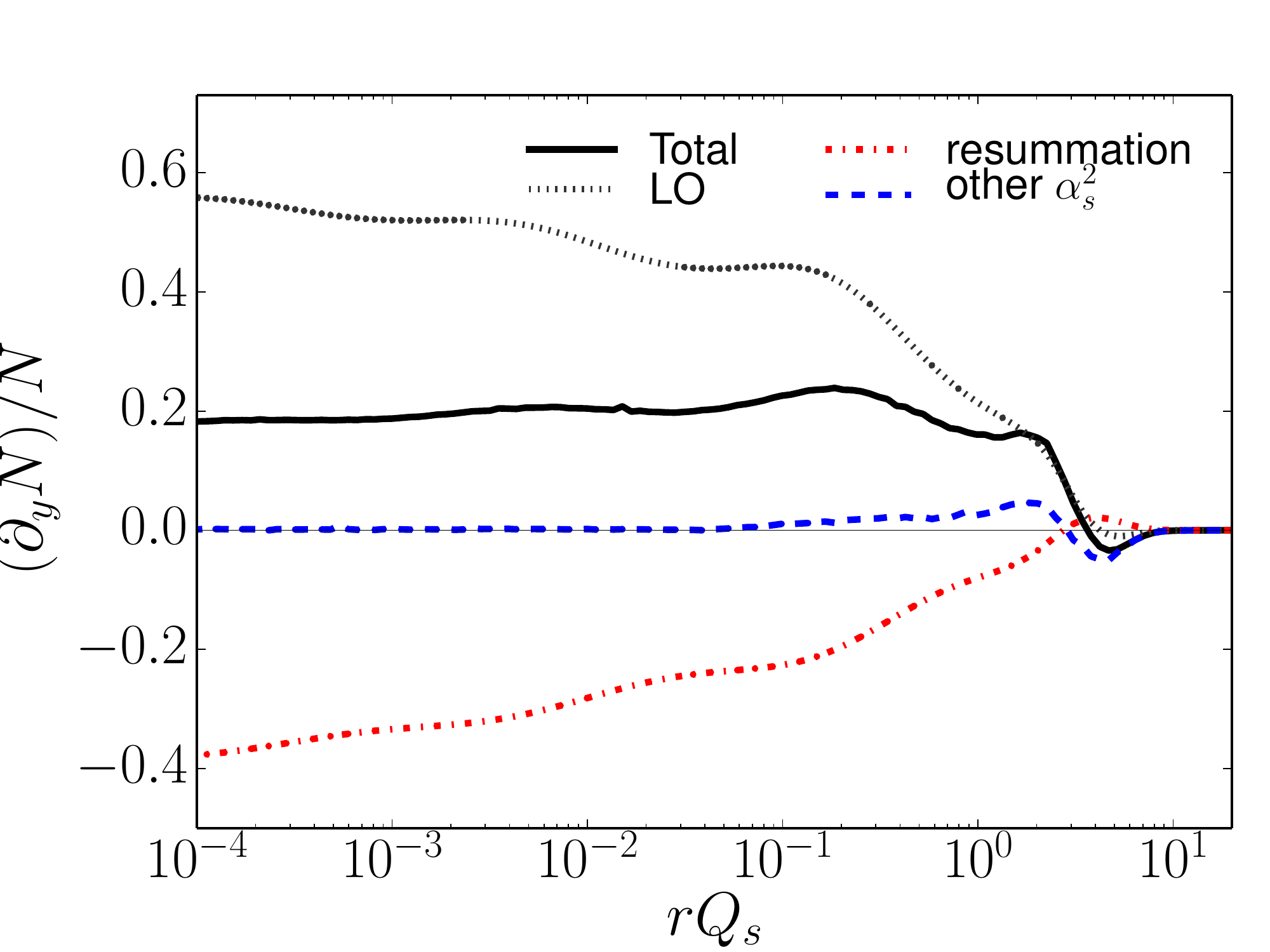} 
\caption{Evolution speed of the dipole amplitude at the initial condition.}
\label{fig:dn_y0} 
\end{minipage}
\hspace{0.5cm}
\begin{minipage}[t]{0.48\linewidth}
\centering
\includegraphics[width=1.05\textwidth]{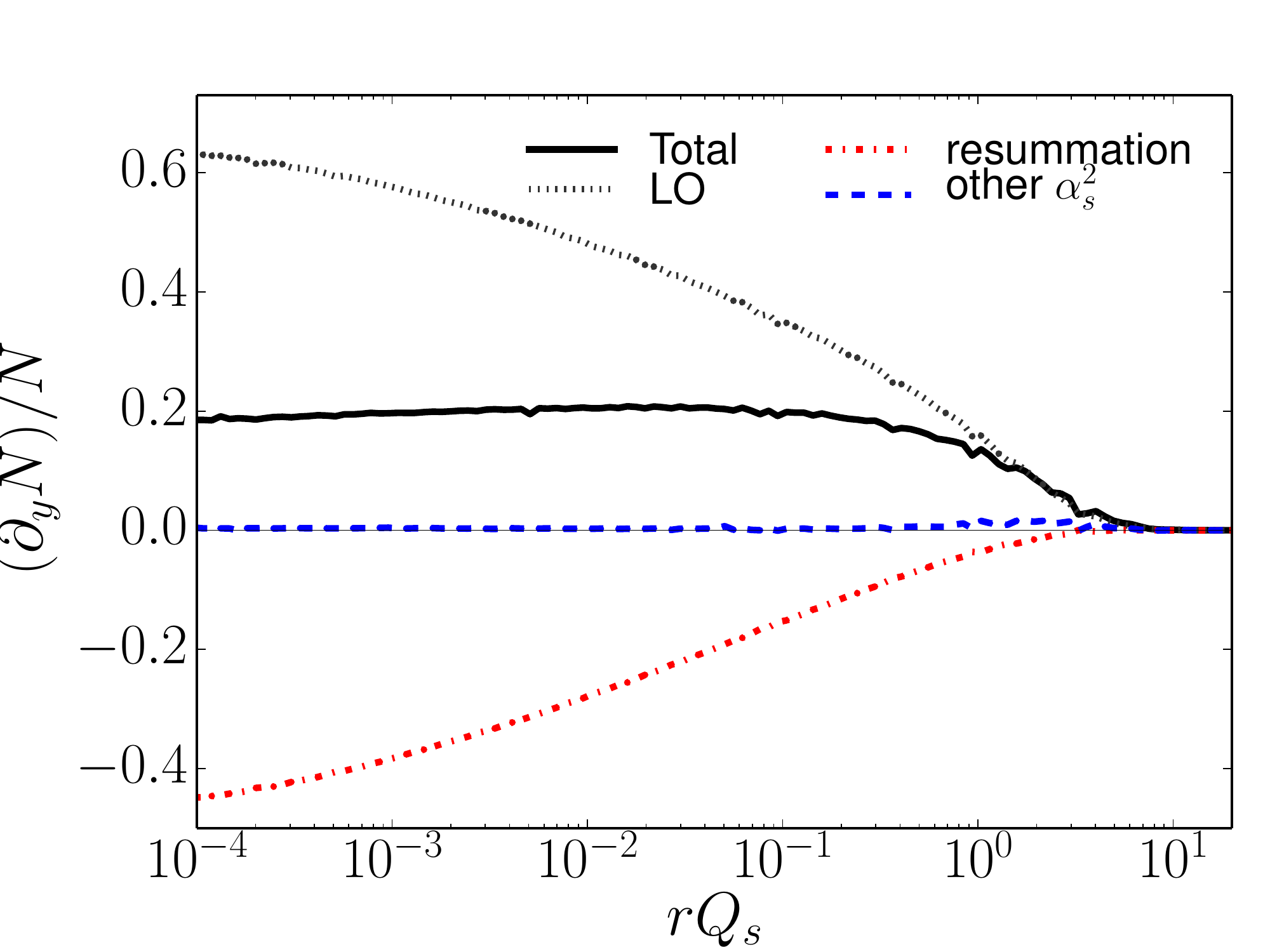} 
\caption{Evolution speed of the dipole amplitude at $y=10$.}
\label{fig:dn_y10} 
\end{minipage}
\end{figure} 

The evolution speed as a function of dipole size is studied in Fig.~\ref{fig:dn_y0} where we show the rapidity derivative of the dipole amplitude as a function of parent dipole size at the initial condition. In addition, different contributions to $(\partial_y N(r))/N(r)$ are shown. The resummation contribution is defined by calculating the contribution from the resummed part of the LO BK kernel $K_\mathrm{Bal}K_\mathrm{DLA} K_\mathrm{STL}$ and subtracting the LO BK contribution with running coupling (for explicit expressions, see Ref.~\cite{Lappi:2016fmu}). The resummation effects can be seen to significantly reduce the evolution speed at small dipoles. 

The fixed order $\as^2$ contribution consists of contributions originating from kernels $K_\mathrm{Sub}$, $K_1^\mathrm{fin}$, $K_2$ and $K_f$ that are not enhanced by large logarithms. These other NLO terms are found to have small positive contribution to the evolution speed at small dipoles, the contribution becoming numerically comparable to the resummation contribution around $r\sim 1/Q_s$.   Note that modifying the value of $C_\mathrm{sub}$ moves contributions between the resummation and the other $\as^2$ terms, and we have checked that it does not significantly affect the overall evolution.

The oscillations seen in Fig.~\ref{fig:dn_y0} originate from resummation of the initial condition. These oscillations are washed away in the evolution, as at larger rapidity $y=10$ they are not visible anymore, see Fig.~\ref{fig:dn_y10}. This can already be seen from Fig.~\ref{fig:amplitude}, where the oscillations present at the initial condition at small dipoles are not visible anymore at $y=5$. At larger rapidities we also find that the fixed order $\as^2$ terms become negligible over the whole range of parent dipole sizes.

\section{Conclusions}
We have shown in Ref.~\cite{Lappi:2016fmu} that when large single and double transverse logarithms are resummed to all orders, the NLO BK equation becomes stable and the problems like negative evolution speed found in Ref.~\cite{Lappi:2015fma} are fixed. By solving the resummed NLO BK equation numerically we find that the evolution speed is significantly reduced compared to the leading order BK equation with running coupling. The $\as^2$ terms that are not enhanced by large transverse logarithms were also found to be numerically important close to phenomenologically relevant initial conditions. We conclude that with resummation contributions included the NLO BK equation can be applied in phenomenological NLO calculatoins.

\section*{Acknowledgements}
We thank D. Triantofyllopoulos and R. Paatelainen for discussions. This work has been supported by the Academy of Finland, projects 267321 and 273464, and by computing resources from CSC – IT Center for Science in Espoo, Finland. H. M. is supported under DOE Contract No. DE-SC0012704.

\bibliographystyle{h-physrev4mod2.bst}
\bibliography{../../refs}

\end{document}